\documentclass[11pt,twoside]{article}

\usepackage{asp2004}
\usepackage{epsf}
\usepackage{psfig}
\usepackage{lscape}

\begin{document}
\title{Circumstellar rings, flat and flaring discs}
\author{M.L. Arias$^1$, J. Zorec$^2$, Y. Fr\'emat$^3$} 
\affil{$^1$ Facultad de Ciencias Astron\'omicas y Geof\'\i sicas, UNLP, 
Argentina}
\affil{$^2$Institut d'Astrophysique de Paris, UMR7095 CNRS, Univ. P\&M Curie}
\affil{$^3$Royal Observatory of Belgium}

\begin{abstract}
 Emission lines formed in the circumstellar envelopes of several type of stars
can be modeled using first principles of line formation. We present simple 
ways of calculating line emission profiles formed in circumstellar envelopes 
having different geometrical configurations. The fit of the observed line 
profiles with the calculated ones may give first order estimates of the 
physical parameters characterizing the line formation regions: opacity, size,
particle density distribution, velocity fields, excitation temperature.
\end{abstract}

\vspace{-0.7cm}

\section{Rings and flaring discs}

\begin{figure}[t]
\centerline{\psfig{file=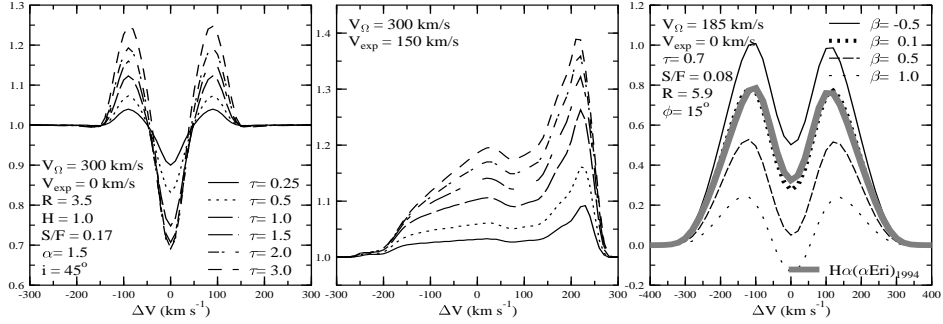,width=12.5truecm,height=4.3truecm}}
\caption[]{a) Line profiles for a ring with $V_{exp}\!=\!0$ and the indicated
parameters. b) 'Steeple' line profiles produced by rings with same parameters
as in a) but $V_{\exp}\!\neq\!0$. c) Line profiles due to flaring discs 
treated as rings, for several $\beta$ and opening angle $\phi\!=\!15^o$. 
$\beta\!=$ 0.1 closely fits H$\alpha$ of $\alpha$~Eri in 1994 (a flat ring 
would require $H\!=\!3.8$; $\beta\!=\!-0.1$)}
\label{f1}
\end{figure}

 In optically thick regions, more than 90\% of the emitted energy in a given 
line can be considered produced in a limited region of the circumstellar disc.
This is the case of Fe\,{\sc ii} in Be stars (Arias et al. 2005), but also 
hydrogen Balmer, Paschen series if their formation region has the aspect of an 
expanding ring. In both circumstances, the emitted radiation can be estimated
using an equivalent isothermal ring with uniform semi-height $H$, whose 
surface density equals the radial column density. If the particle density has
a distribution $N(R)\!\sim$ $R^{-\beta}$, the radius of the ring is then 
$R_r/R_*\!=\!$ $[(1-\beta)/(2-\beta)][1-(R_i/R_e)^{\beta
-2}]/[1-(R_i/R_e)^{\beta-1}]$, where $R_{i,e}$ are the internal and external 
radii of the line forming region. The ring can be considered having expansion
$V_{exp}$ and rotation $V_{\Omega}$ velocities, both represent averages of the 
respective velocity fields (Arias 2004). The emitted flux is simply 
$F_{\lambda}=$ $\int_{{\cal S}(i)}I_{\lambda}(x,y){\rm d}x{\rm d}y$ where 
${\cal S}(i)$ is the aspect-angle effective emitting surface projected on the
sky and $I_{\lambda}(x,y)$~is:
\begin{equation}
\left.\begin{array}{lcl}
I^a(x,y,v-v_r) & = & I^*(x,y)e^{-{\tau^f(v-v_r)\over\mu(x,y)}}+
S[1-e^{-{\tau^f(v-v_r)\over\mu(x,y)}}] \\
 &  & ({\rm stellar\ emission\ absorbed\ by\ the\ front\ side}) +\cr
 &  & ({\rm front\ side\ emission})\\
I^b(x,y,v-v_r) & = & S[1-e^{-{\tau^t(v-v_r)\over\mu(x,y)
                 }}]e^{-{\tau^f(v-v_r)\over\mu(x,y)}}+
                 S[1-e^{-{\tau^f(v-v_r)\over\mu(x,y)}}] \\
 &  & ({\rm rear\ emission\ absorbed\ by\ the\ front\ side}) +\\
 &  & ({\rm front\ side\ emission}) \\
\end{array}
\right\}
\end{equation}
\noindent with $\tau_v\!=$ $\tau_o\Phi(v-v_r)$ and $v_r(x,y)\!=$ $\pm
\{V_{exp}(R)[1-({x\over R})^2]^{1\over2}\pm V_{\Omega}(R){x\over R}\}\sin i$,
where $(f,r)$ stand for `front' and `rear' sides of the ring. The signs in the 
radial velocity $v_r$ are chosen according to the quadrant; $v$ is the Doppler 
displacement in the observed emission line profile, $\mu(x,y)\!=$
$\cos$(ring-normal;observer), $i$ inclination of the ring-star system, 
$\tau_o$ is the radial opacity of the ring in the line center and $\Phi(v)$ is 
the intrinsic absorption line profile. The line source function is: 
$S_{\lambda}(\tau_{\rm o})\!=$ $S_o$ for $\tau_o\leq1$; $S_o\tau_{\rm o}^p$ 
for $\tau_o\!>\!1$ [$p\!\simeq\!1/2$ for Gaussian $\Phi(v)$] and 
$B_{\lambda_o}(T_{\rm e})$ for $\tau_o\!\geq$ $(B_{\lambda_o}/S_o)^2$ (Mihalas 
1978, Chap.11). $S_o$ depends on the nature of the line transition: collision-, 
photoionization-, mixed-dominated. Using $S_o$ as done in Cidale \& Ringuelet
(1989) we can determine the excitation temperature (Arias et al. 2005). Thus,
the free parameters to fit the observed emission line profiles are: $S_o/F_*$,
$\tau_o$, $R_r$, $H$, $i$, $V_{\Omega}$, $V_{exp}$ and density distribution 
$\beta$ (or $\alpha\!=$ $2-\beta$). The shape of the line profiles reduce 
strongly the space of free parameters, mainly those of the velocity field and
the ratio between $\tau_o$ and $H$. Line intensities are sensitive to 
$(S_o/F_*)R^2_r$ and $R_r$ to the temperature, which in turn determines 
$S_o/F_*$. Figure~\ref{f1}a shows line profiles obtained for the indicated 
parameters of the ring with $V_{exp}\!=\!0$, which are typical for {\it 
`shell'} lines. The same ring parameters are used for Fig.~\ref{f1}b, but 
$V_{exp}\!=$ 150 km/s. The line profiles are of the {\it`steeple'} type seen 
frequently in Fe\,{\sc ii} lines. Flaring discs can be treated in the same way 
as cylindrical ones, except that the surface density, and hence $\tau_o$ 
depends on the coordinate perpendicular to the equator (Vinicius et al. 2005). 
A fit of the H$\alpha$ line of $\alpha$~Eri in 1994 with a flaring disc of 
opening angle $\phi=15^o$ is shown in Fig.~\ref{f1}c, where are also shown 
line profiles for different $\beta$-values. 
\begin{figure}[]
\centerline{\psfig{file=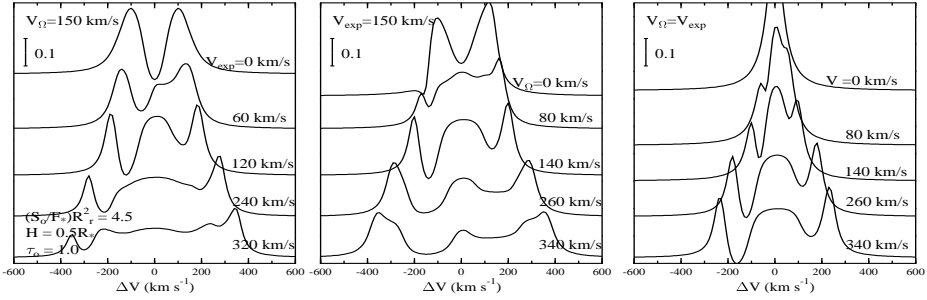,width=12.5truecm,height=4truecm}}
\caption[]{Three-peak line profiles produced by rotating+expanding rings}
\label{f2}
\end{figure}
Several three-peak emission line 
profile are shown in Fig.~\ref{f2} produced in regions treated as equivalent 
expanding rings. Due to the Doppler shifts, the central emission components 
are produced by the front and rear sides of ring sectors where the radial 
velocity is $v_r\simeq0$.
\section{Discs}

\begin{figure}[]
\centerline{\psfig{file=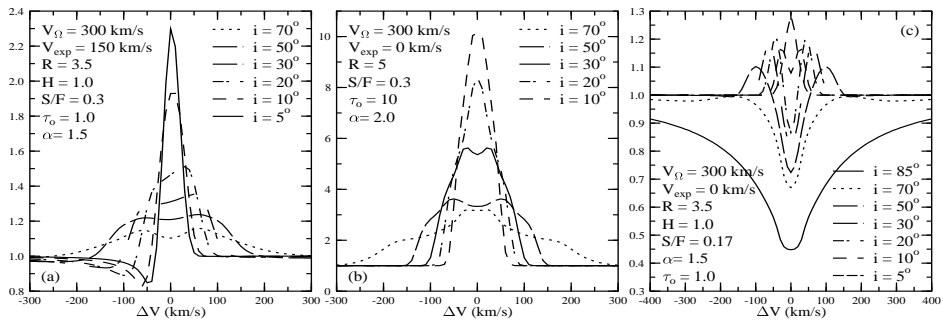,width=12.5truecm,height=4.3truecm}}
\caption[]{Emission line profiles obtained with discs. a) P Cyg like profiles.
b) Bottle shaped. c) Broadening due to $v_1$ at $i\to90^o$}
\label{f3}
\end{figure}

 The main difference in the treatment of discs with respect to that for rings
is in the velocity dependence of the intrinsic absorption line profile. It has 
been shown by Horne \& Marsh (1986) that for a Gaussian $\Phi(v)$ the Doppler
width of the profile is enlarged by a ``turbulent" term due to the differential 
rotation in the disc towards the observer's direction. The wavelength 
dependent opacity is then proportional to $exp\{-(1/2)[(\lambda-\lambda
_D)/(\Delta_D\times\delta)]^2\}$, where $\lambda_D\!=$ $\lambda_o(v_o/c)$~and:
\begin{equation}
\left.\begin{array}{lcl}
v_o & = & [V_{\Omega}(R)\sin\theta+V_{exp}(R)\cos\theta]\sin i \\
\delta & = & [1+(\lambda_ov_1/c\Delta_D)^2]^{1/2} \\
v_1 & = & [{1\over2}V_{\Omega}(R)\sin\theta+V_{exp}(R)(2-
\beta)\cos\theta]\cos\theta\tan i\sin i \\ 
\end{array}
\right\}
\end{equation}
\noindent where $\theta$ is the azimuthal angle. For large values of $H$, 
$v_1$ acts as a non-negligible broadening agent of the effective Doppler line
width. Different examples of line profiles obtained with discs seen at several
inclination angles are shown in Fig~\ref{f3}. The parameters used in 
Fig~\ref{f3}a can suite for P~Cyg type line profiles, while Fig~\ref{f3}b 
reproduce the bottle shaped emission line profiles seen frequently in Be stars. 
The {\it `bottle'} shaped profiles can be obtained also with rings. They are 
produced by the $\tau_o^p$ opacity dependence of the source function due to 
its non-local energy supply. The broadening of the effective Doppler line by 
$v_1$ is depicted in Fig~\ref{f3}c ($i\!=\!85^o$). Other related subjects can
be found in http://www2.iap.fr/users/zorec/.
 
\end{document}